\documentclass[pra,onecolumn,secnumarabic,superscriptaddress,showpacs,
showkeys,amssymb,amsmath,nobibnotes,aps]{revtex4}

\usepackage{graphics}              
\usepackage{latexsym}
\usepackage{bm}                    
\usepackage{amsthm,amsfonts,url}   
\usepackage{verbatim}              

\expandafter\ifx\csname package@font\endcsname\relax\else
 \expandafter\expandafter
 \expandafter\usepackage
 \expandafter\expandafter
 \expandafter{\csname package@font\endcsname}
\fi

\def\blfootnote{\xdef\@thefnmark{}\footnotetext}

\begin{document}

\title{Exact interferometers for the concurrence and residual $3$-tangle}


\author{Hilary A. Carteret}
\affiliation{Department of Combinatorics and Optimization,
             University of Waterloo, Waterloo, Ontario, N2L 3G1, Canada}

\date{September 4, 2006}

\begin{abstract}
In this paper we describe a set of circuits that can measure the concurrence 
of a two qubit density matrix without requiring the deliberate addition 
of noise.  We then extend these methods to obtain a circuit to measure 
one type of three qubit entanglement for pure states, namely the $3$-tangle.  
\end{abstract}

\maketitle


\section{Introduction}

\begin{figure}[ht!]
    \begin{minipage}{\columnwidth}
    \begin{center}
        \resizebox{0.6\columnwidth}{!}{\includegraphics{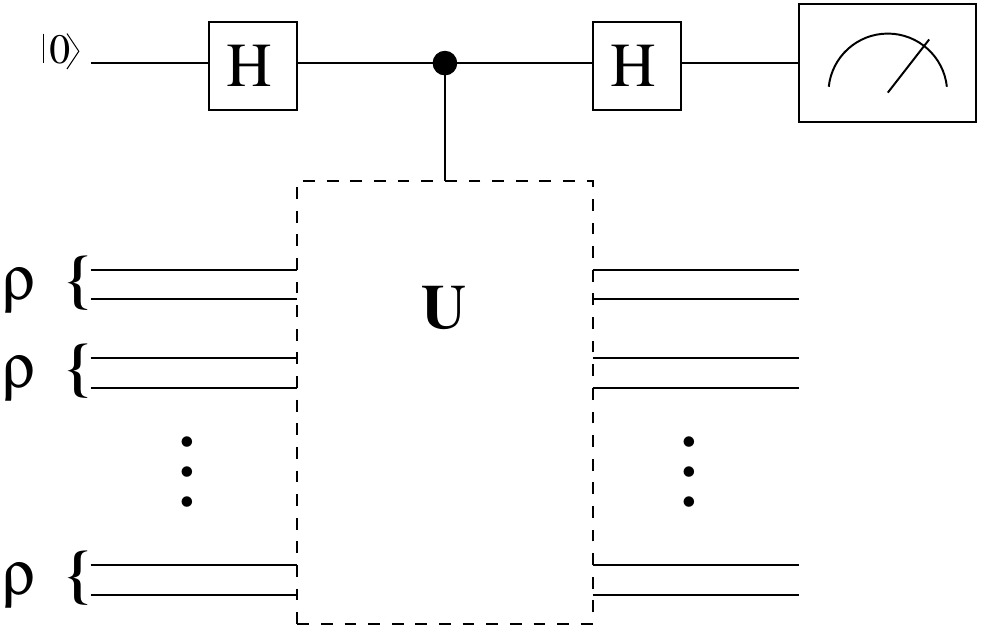}}
    \end{center}
    \end{minipage}
    \caption{General interferometer circuit}
    \label{interferometer}
\end{figure}

Techniques for measuring entanglement without first reconstructing the state 
have recently attracted a lot of attention.  Some of these approaches have 
been based on the Structural Physical Approximation (SPA) \cite{SPA}, 
followed by measuring the spectrum of the resulting density matrix 
\cite{Direct,DirectNLF,LLW}
(but see also \cite{Toddstuff,Filip1}).
One method for measuring that spectrum was inspired by \cite{GPmixed}, 
in which it was shown that while the physical evolution of the density 
matrix must be $\rho \mapsto U\rho U^{\dagger},$ the interference pattern 
is proportional to Re$(ve^{-i\varphi}) = $Re(Tr$(U\rho)),$ where $v$ is the 
visibility and $\varphi$ is the phase shift.  Thus the interferometer 
circuit in Figure~\ref{interferometer} 
can modify the expectation value of the measured qubit, which corresponds 
to the path in a Mach-Zehnder interferometer.  
Note that $U$ can be any unitary operation. 

In \cite{NoiselessPT} we showed how to measure the spectrum of 
$\rho^{{\rm{T}}_2}$ without using the SPA to shift the spectrum to be 
non-negative.  The ``Wootters spin-flip'' used to define the concurrence is 
also an unphysical operation.  The techniques in \cite{NoiselessPT} cannot 
be used directly to measure this invariant, but a more general type 
of circuit can. 

We can measure the $3$-tangle \cite{3tangle} in a similar way, but some 
of the algebra used in the design of these circuits is only valid for 
pure states; the function conflates subsystem mixing with entanglement, 
and so these circuits will have the same restrictions on their validity.

\section{A Generalized Interferometer}

We will begin by generalizing the circuit construction in \cite{GPmixed}. 
This studied the behaviour of a Mach-Zehnder interferometer. 
By assumption, our particle has an internal degree of freedom, or spin.  
The circuit couples the internal and external degrees of freedom, 
by the operation
\begin{equation}\label{Uint}
 {\bm{U}}=
 \begin{pmatrix} 0 & 0 \\
                 0 & 1 
 \end{pmatrix}
 \otimes 
 U^{(i)} + 
 \begin{pmatrix} e^{i\chi} & 0 \\
                 0         & 0 
 \end{pmatrix}
 \otimes \openone^{(i)}.
\end{equation}
When the control qubit is ``on'', the internal degrees of freedom $\rho_0$ 
undergo the evolution
 $\rho_0 \mapsto U^{(i)} \rho_0 U^{(i)\dagger}$
where the label $(i)$ signifies that only the internal degrees of freedom 
are affected.  Note that $U^{(i)}$ can be any unitary matrix that only 
acts on the internal degrees of freedom.  A little algebra then gives us 
that the output intensity,
\begin{equation}\label{intensity}
 I \propto 1+|{\rm{Tr}}(U^{(i)}\rho_0)|\cos[\chi-\arg{\rm{Tr}}(U^{(i)}\rho_0)].
\end{equation}
We will now modify this circuit to generalise \eqref{Uint} to a completely 
positive (CP) map.  (While this question has been considered before in 
\cite{GPCPmaps}, this had a different motivation and was not concerned with 
measuring entanglement.)  We will limit ourselves to convex combinations of 
unitary operators, and introduce a family of unitary transformations 
labelled by the index $k,$ 
\begin{equation}\label{Unitalint}
 {\bm{U}}_k=
 \begin{pmatrix} 0 & 0 \\
                 0 & 1 
 \end{pmatrix}
 \otimes 
 U_k^{(i)} + 
 \begin{pmatrix} e^{i\chi} & 0 \\
                 0         & 0 
 \end{pmatrix}
 \otimes \openone^{(i)}.
\end{equation}
ote that these operators differ only in the unitary operator $U^{(i)}_k$ 
acting on the internal degrees of freedom.  If we weight these with  
probabilities $p_k,$ then the controlled evolution of the internal degree 
of freedom is now 
\begin{equation}
 \rho_0 \mapsto \sum_k p_k U_k^{(i)} \rho_0 U_k^{(i)\dagger}.
\end{equation}
When we redo the calculation in \cite{GPmixed} with our generalized internal 
dynamics, we obtain the measured intensity along the $|0\rangle\langle 0|$ 
arm of the interferometer, which will be proportional to 
\begin{equation}
 1+\left|{\rm{Tr}}\left(\sum_k p_k U_k^{(i)}\rho_0\right)\right|
    \cos\left[\chi-\arg{\rm{Tr}}\left(\sum_k p_k U_k^{(i)}\rho_0\right)\right].
\end{equation}
We will now use this to measure the concurrence.

\section{Measuring the Concurrence}

The concurrence \cite{concurrence} is a bipartite entanglement measure 
defined on mixed states of two qubits to be  
\begin{equation}\label{Cdef}
 C_{AB} = \max\{\lambda_1-\lambda_2-\lambda_3-\lambda_4,0\}
\end{equation}
where the $\lambda_i$s are, in decreasing order of magnitude, the positive 
square roots of the eigenvalues of $\rho_{AB}\widetilde{\rho_{AB}},$ where
\begin{equation}\label{tildadef}
 \widetilde{\rho_{AB}} = \sigma_y\otimes\sigma_y \rho_{AB}^{\rm{T}} 
                         \sigma_y\otimes\sigma_y 
\end{equation}
and where $\;^{\rm{T}}$ denotes matrix transposition in the computational 
basis. The method in \cite{NoiselessPT} will not work, because we will 
obtain acausal circuits in which some of the input rails double back on each 
other.  If we had wanted the spectrum of $\widetilde{\rho_{AB}}$ on its own, 
we could measure that in the same way as in \cite{NoiselessPT}.  However, 
when we try to pre-multiply by $\rho_{AB},$ we find that we have to sum 
two pairs of input indices with two other input indices and likewise with 
two pairs of output indices.  The SPA \cite{withoutprior} would also have 
serious problems because of that matrix multiplication,
as it will instead measure the spectrum of 
\begin{equation}
  \rho_{AB}\Lambda(\rho_{AB})=\mu \rho_{AB} 
  + \nu \rho_{AB}\sigma_y\otimes\sigma_y
      \rho_{AB}^{\rm{T}}\sigma_y\otimes\sigma_y,
\end{equation} 
where $\mu >> \nu$ are the parameters defining the SPA for this map. 
In effect, the SPA circuits will measure the eigenspectrum of $\rho_{AB},$ 
with a small correction term proportional to $\rho_{AB}\tilde{\rho_{AB}}.$  
Without prior knowledge of $\rho_{AB}$ there is no way to correct this error.

We will now describe two alternative methods for measuring the concurrence.  
Recall that 
\begin{equation}
 \sigma_y =
 \begin{pmatrix} 0 & i \\
                -i & 0 
 \end{pmatrix}
 = i \varepsilon, \quad \text{where} \quad \varepsilon = 
 \begin{pmatrix} 0 & 1 \\
                -1 & 0
 \end{pmatrix}.
\end{equation}
Rewriting equation \eqref{tildadef} in index notation and using the Einstein 
summation convention, we obtain
\begin{equation}
 \widetilde{\rho}_{ij}^{k\ell}=(\varepsilon\otimes\varepsilon)_{ij}^{rs}
 (\rho^{{\rm{T}}})_{rs}^{pq}(\varepsilon\otimes\varepsilon)_{pq}^{k\ell},
\end{equation}
where we have omitted the $AB$ subscripts to reduce clutter.  Converting the 
$\varepsilon$ matrices to vectors, we can write
\begin{equation}
 \widetilde{\rho}_{ij}^{k\ell}=\varepsilon_{ir}\varepsilon_{js}
                               \varepsilon^{pk}\varepsilon^{q\ell}
                               \rho_{pq}^{rs}.
\end{equation}
Now use the identity 
 $\varepsilon_{ab}\varepsilon^{cd}=\delta_a^c\delta_b^d-\delta_a^d\delta_b^c$
to obtain \cite{3qubits}
\begin{equation}
 \widetilde{\rho}_{ij}^{k\ell} = \rho_{ij}^{k\ell} 
                                 - \rho_{rj}^{r\ell}\delta_i^k 
  - \delta_j^{\ell}\rho_{is}^{ks} + \delta_i^k\delta_j^{\ell}\rho_{rs}^{rs}.  
\end{equation}
In the more familiar matrix notation, this is
\begin{equation}
 \widetilde{\rho_{AB}} = \rho_{AB} - \openone_{A} \otimes \rho_{B} 
           -\rho_{A} \otimes \openone_{B} + \openone_{AB}.
\end{equation}
We can now safely perform the matrix multiplication:
\begin{equation}\label{sumform}
 \rho_{AB}\widetilde{\rho_{AB}}=\rho_{AB}^2 
                                - \rho_{AB}(\openone_{A} \otimes \rho_{B})
                                - \rho_{AB}(\rho_{A}\otimes \openone_{B})
                                + \rho_{AB}.
\end{equation}
We will need to find a way to measure the various moments of 
$\rho_{AB}\widetilde{\rho_{AB}}$ to determine the eigenspectrum and thus 
obtain the $\lambda_i$s for \eqref{Cdef}. 
There are two ways in which we can convert equation \eqref{sumform} into an 
interferometer. 
The first method uses the modified interferometry circuit introduced in 
the previous section directly, as it enables us to measure the effects 
of sums of unitary matrices, despite the fact that no physical evolution 
can be written as the sum of unitary matrices in this way.   
It should be noted that we will be needing the moments of $\rho\tilde{\rho},$
rather than $\tilde{\rho}$ itself.  So, the fact that some of the terms in 
$\tilde{\rho}$ do not have trace $1$ will not require us to rescale those 
terms, as $\rho \openone = \rho,$ so the second set of rails can just be 
ignored, without needing to find a way to change the relative weights of 
the terms. 

\begin{figure}[floatfix]
    \begin{minipage}{\columnwidth}
    \begin{center}
        \resizebox{0.7\columnwidth}{!}{\includegraphics{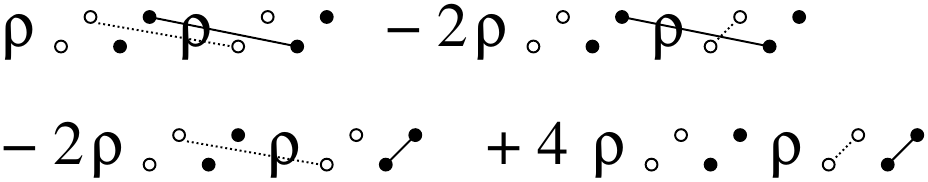}}
    \end{center}
    \end{minipage}
    \caption{Summation diagram for $\rho_{AB}\widetilde{\rho_{AB}.}$}
    \label{4sum}
\end{figure}

The permutations required for each term can be decomposed into two operations 
which are variously ``on'' or ``off'' for each term in \eqref{sumform}. 
The set of probabilities also factorizes conveniently, so we will need only 
two auxiliary control ancillas.  
Figure~\ref{4sum} shows the terms of \eqref{sumform} as a summation diagram, 
which is a pictorial version of index notation.  The nodes represent indices 
and the links represent summations.  Unlinked nodes are free indices.  The 
nodes on each copy of $\rho$ are ordered as for standard tensor index 
notation.   
The summation pattern for the last term is defined to have both ancillas in 
the ``off'' position.  The corresponding circuit diagram for this 
{\emph{single}} term could be obtained by rotating the Figure clockwise by 
$-90^{\circ},$ omitting the $\rho$ symbols and inserting the resulting 
wiring pattern into the box marked ``$U$'' in Figure~\ref{interferometer}.  
In this case, all the rails go straight through and this sub-circuit would 
measure the trace norm if implemented on its own.  Contrast this with the 
first term which will have both ancillas ``on'' and is proportional to 
Tr$(\rho_{AB}^2).$  Note that the only difference between these two terms 
is a pair of swaps.

\begin{figure}[ht!]
    \begin{minipage}{\columnwidth}
    \begin{center}
        \resizebox{0.6\columnwidth}{!}{\includegraphics{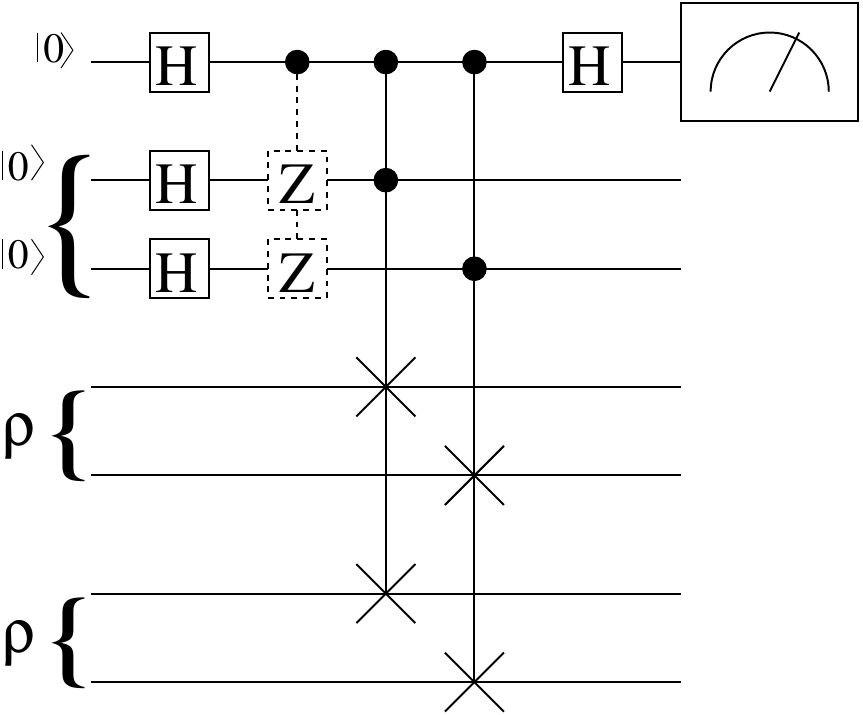}}
    \end{center}
    \end{minipage}
    \caption{The circuit for $\tfrac{1}{4}$Tr$\rho_{AB}\widetilde{\rho_{AB}}.$}
    \label{circuit1}
\end{figure}

We must also take care to ensure that we reproduce the minus signs in 
equation \eqref{sumform}, so the qubit we are going to measure will 
be able to ``see'' the minus signs from the ancillas. 
This can be done either by initializing the control ancillas with the 
appropiate relative phase, together with an initial controlled-$\sigma_Z$ 
gate, or by including the desired phase in the gate itself \cite{GangOf9}.  
We will use the first method and we will therefore need to initialize both 
auxiliary ancillas in the state 
$|-\rangle = \tfrac{1}{\sqrt{2}}(|0\rangle - |1\rangle).$
The fact that the ancillas must be normalized means that each auxiliary 
ancilla will decrease the visibility of the interference fringes by a 
factor of $2.$
If we now insert these components into a modified version of 
Figure~\ref{interferometer} so they are controlled by the two auxiliary 
ancillas, we will obtain the full circuit for 
Tr$(\rho_{AB}\widetilde{\rho_{AB}})$ in Figure~\ref{circuit1}.

\begin{figure}[ht!]
    \begin{minipage}{\columnwidth}
    \begin{center}
        \resizebox{0.5\columnwidth}{!}{\includegraphics{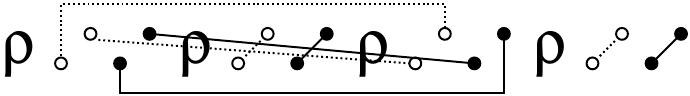}}
    \end{center}
    \end{minipage}
    \caption{All auxiliary control ancillas off.}
    \label{4off}
\end{figure}

\begin{figure}[ht!]
    \begin{minipage}{\columnwidth}
    \begin{center}
        \resizebox{0.5\columnwidth}{!}{\includegraphics{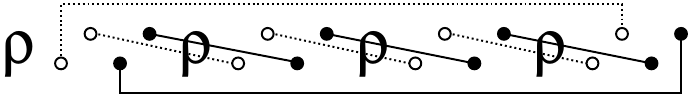}}
    \end{center}
    \end{minipage}
    \caption{All auxiliary control ancillas on.}
    \label{4on}
\end{figure}

\begin{figure}[ht!]
    \begin{minipage}{\columnwidth}
    \begin{center}
        \resizebox{0.8\columnwidth}{!}{\includegraphics{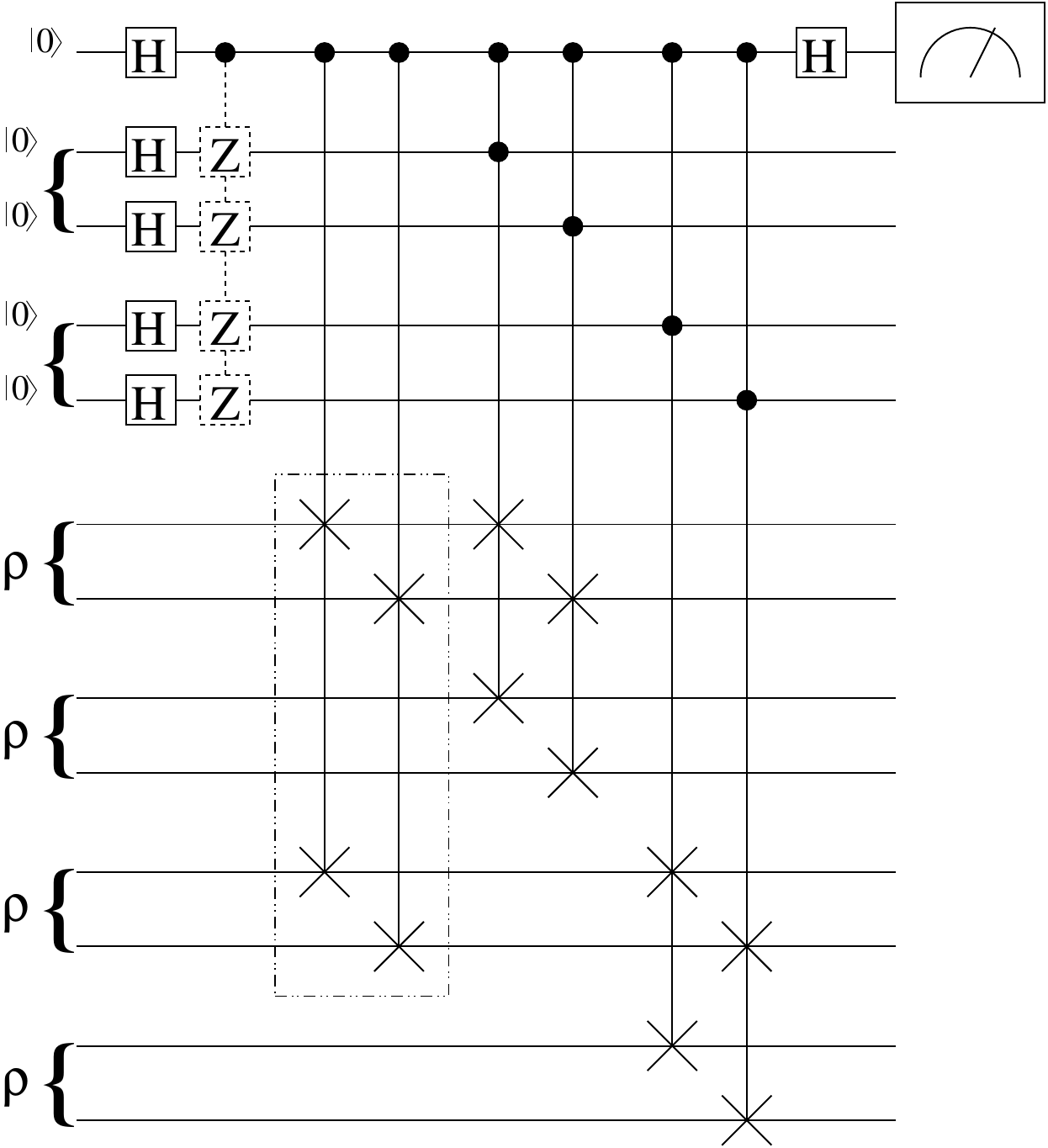}}
    \end{center}
    \end{minipage}
    \caption{The circuit for $\tfrac{1}{16}$Tr$((\rho\widetilde{\rho})^2).$}
    \label{circuit2}
\end{figure}

The same principles can be used to construct the circuit for  
Tr$((\rho_{AB}\widetilde{\rho_{AB}})^2).$  This has 16 terms and will require 
four auxiliary control ancillas, initialized in the same way.  The structure 
of this circuit can be parsed as follows.  The box in dotted lines with no 
auxiliary controls performs the permutation for the matrix multiplication 
of the two copies of $\rho_{AB}\widetilde{\rho_{AB}}.$  The term with all the 
auxiliary control ancillas off will look like Figure~\ref{4off}.
\newline
Figure~\ref{4on} is the diagram for the term with all the ancilla-controlled 
operations ``on''.   Again, the difference between the two is just a set of 
simple swaps, 
\newline
which can be controlled by auxiliary ancillas as before.  Thus we obtain 
the circuit for measuring $\tfrac{1}{16}$Tr$((\rho\widetilde{\rho})^2),$ 
in Figure~\ref{circuit2}.
The circuits for $\tfrac{1}{64}$Tr$((\rho_{AB}\widetilde{\rho_{AB}})^3)$ and 
$\tfrac{1}{256}$Tr$((\rho_{AB}\widetilde{\rho_{AB}})^4)$ can be constructed 
in a similar way.  We can now calculate the spectrum $\{\eta_i\}$ of 
$\rho_{AB}\widetilde{\rho_{AB}}$ using the recipe given in \cite{Direct,KeylW}.
The positive square roots of the $\eta_i$s give the concurrence spectrum 
and equation \eqref{Cdef} can then be used to compute the concurrence.  

This first method shows that the concurrence (and hence the entanglement of 
formation) can be measured without any prior knowledge of the state.  However,
this method is not very sensitive, because the visibility drops by a factor of 
$\tfrac{1}{4^m}$ for the $m$th order concurrence moment.  Fortunately, 
there is a second method for measuring the moments of the concurrence 
spectrum.  This uses the principle of implicit measurement \cite{implicit}, 
whereby we can assume the auxiliary ancillas have been measured at the end 
of the circuit without affecting the statistics of the principal
measurement.  These circuits are examples of ``non-erasing quantum erasers'' 
\cite{nonerasure},
albiet very simple ones.  In effect we will be measuring each term in the 
various summations separately.  If we choose to measure the ancillas, we 
will be partitioning the output of the primary ancilla into subensembles 
with greatly enhanced visibility.  We can also simplify the circuits by 
initializing the auxiliary ancillas using only a simple Hadamard gate, 
and omitting the controlled-$\sigma_Z$ gates altogether.

Alternatively, we can implement the circuits for each term separately, and 
insert the necessary weight factors by hand.  This has the additional 
advantage that we can exploit the cyclicity of the trace and the fact 
that Tr$\rho_{AB}=1$ and its first four moments completely define its 
spectrum, to reduce the number of circuits required from $340$ to no more 
than $111.$  This does not compare well to full state tomography (which 
needs only 16 parameters) but this upper bound is unlikely to be tight, 
as this system can be characterised by 18 non-local parameters \cite{makhlin}, 
of which only 9 are fully algebraically independent; the other 9 are  
discrete valued.

\section{Measuring the $3$-tangle}

The residual $3$-tangle as defined in \cite{3tangle} is the modulus of a 
complex number.  It is only defined in closed form for {\emph{pure}} states 
of three qubits.  If we define the $2$-tangle $\tau_{AB}$ to be the square 
of the concurrence \cite{3tangle}, then the residual $3$-tangle is defined 
as entanglement between the qubits that cannot be accounted for in terms 
of bipartite entanglement:
\begin{equation}\label{tau3def}
 \tau_{ABC}=\tau_{A(BC)}-\tau_{AB}-\tau_{AC}.
\end{equation}
The measure is symmetric under relabelling of the three parties \cite{3tangle}.
We cannot use \eqref{tau3def} directly, because we cannot define the 
equivalent of $\sigma_y$ on the merged party $(BC)$ without knowing the 
support of $\rho_{BC},$ which would require detailed knowledge of the state.

The fact that the two-party density matrices are all rank 2 means that 
$\lambda_3=\lambda_4=0.$  So we can write
\begin{equation}
 \tau_{AC} = (\lambda_1-\lambda_2)^2
           = \lambda_1^2-\lambda_2^2-2\lambda_1\lambda_2.
\end{equation}
Wootters {\emph{et al.}} then obtain that 
$\tau_{A(BC)} = 4 \det \rho_A.$
This, together with the definition in \eqref{tau3def} gives us that
\begin{equation}
 \tau_{AB}+\tau_{AC} = 4 \det \rho_A - \tau_{ABC}. 
\end{equation}
Using the relabelling symmetry and the fact that 
det$\rho_i = \tfrac{1}{2}(1-\text{Tr}(\rho_i^2))$ for $2 \times 2$ matrices, 
we can write \cite{Sudbery3}
\begin{equation}\label{taufinal}
 \tau_{ABC} = 2(1 - \text{Tr}(\rho_A^2) - \text{Tr}(\rho_B^2) 
               + \text{Tr}(\rho_C^2) - \tau_{AB}). 
\end{equation}
We can calculate this from the rescaled expectation values measured by the 
circuits for $\tau_{AB}$ we obtained earlier, together with the statistics 
from the circuits for the Tr$(\rho_i^2)$s.  These one-party circuits need two 
copies of their single particle $\rho$s each and the controlled-$U$ operation 
for these is the SWAP gate \cite{Direct}.  We can combine these three 
one-party circuits into a single three party circuit with three control 
qubits, thus saving four copies of $\rho_{123}$ per measurement.
We can therefore measure the $3$-tangle using this family of five circuits, 
provided the state is pure. 

We can improve both the cost and sensitivity for the $3$-tangle as 
follows.  Recall that \cite{3tangle} 
\begin{equation}
 {\rm{Tr}}(\rho_{AB}\widetilde{\rho_{AB}}) + 
 {\rm{Tr}}(\rho_{AC}\widetilde{\rho_{AC}}) = 4 \det \rho_A,
\end{equation}
where the superscripts refer to which two-party density matrix was used to 
obtain the concurrence spectrum.  
We know that 
$\lambda^{(AB)}_1\lambda^{(AB)}_2 
= \lambda^{(BC)}_1\lambda^{(BC)}_2 = \lambda^{(AC)}_1\lambda^{(AC)}_2$
from the relabelling symmetry, so we can omit the superscripts and write
$\tau_{ABC} = 4 \lambda_1\lambda_2.$ 
The square root in the definition of the $\lambda_i$s prevents us from 
measuring this but we {\emph{can}} construct a circuit for 
$|\tau_{ABC}|^2 = 16 \lambda_1^2\lambda_2^2.$  Now we just need to notice that
$ 2\lambda_1^2\lambda_2^2 = (\lambda_1^2+\lambda_2^2)^2
                             -\lambda_1^4-\lambda_2^4,$ 
and then we can write
\begin{equation}\label{tracediff}
 |\tau_{ABC}|^2 = 8\left({\rm{Tr}}(\rho_{AB}\widetilde{\rho_{AB}})^2 - 
  {\rm{Tr}}((\rho_{AB}\widetilde{\rho_{AB}})^2)\right).
\end{equation}
A circuit with visibility $\propto |\tau_{ABC}|^2$ can be obtained 
from Figure~\ref{circuit2}.  We need an equal superposition of the 
sub-circuits with the matrix multiplication (the faint dotted box) switched 
on and off so we will insert a new ancilla in the state $|-\rangle$ and 
another controlled-$\sigma_Z$ as we want the term with the matrix 
multiplication on to acquire a minus sign.  The intensity along the 
$|0\rangle\langle 0|$ arm for this na{\"\i}ve circuit will be \cite{GPmixed}
\begin{equation}
 I_{|\tau_{ABC}|^2} \propto 1 + |\tau_{ABC}|^2/32.
\end{equation}
This circuit needs four copies of the state per run, compared with the first 
method which needs 22 copies.  The sensitivity of this na{\"\i}ve circuit 
can also be improved by measuring the terms separately.  This can in fact 
be done with just 14 circuits, compared with the 16 parameters that must 
be determined to learn the state of a three qubit state that is known 
to be pure \cite{cavesandco}.

\section{Discussion}

We have presented circuits for both the concurrence and the residual 
$3$-tangle.  The circuits for the concurrence should work for any two qubit 
state.  The visibility of these circuits is not much better than that 
for the SPA circuits, however, the former do not introduce state-dependent 
errors.  The visibility can be improved by decomposing the original moment 
circuits into unitary circuits, but there will then be many more of them. 

The two methods for finding the residual $3$-tangle are, strictly speaking, 
only valid for pure states.  In any laboratory setting the states will have 
a certain amount of mixing.  The effect of this on the residual 
$3$-tangle is incompletely understood so it is unclear how useful 
these circuits would be in practice, but at least these circuits do not 
aggravate the mixing beyond that caused by unavoidable experimental noise.  
They also require fewer circuits than the corresponding state tomography 
problem.


\begin{acknowledgments}

I would like to thank Todd Brun, Martin R{\"{o}}tteler and Michele Mosca 
for useful feedback on this manuscript, and Stephen Bullock and Dianne 
O'Leary for spotting an error in the circuit diagrams in an earlier 
version of this paper. 
This research was supported by  MITACS, The Fields Institute, 
and the NSERC CRO project ``Quantum Information and Algorithms.''

\end{acknowledgments}





\providecommand{\bysame}{\leavevmode\hbox to3em{\hrulefill}\thinspace}
\providecommand{\MR}{\relax\ifhmode\unskip\space\fi MR }
\providecommand{\MRhref}[2]{%
  \href{http://www.ams.org/mathscinet-getitem?mr=#1}{#2}
}
\providecommand{\href}[2]{#2}

\end{document}